\documentclass[12pt,letterpaper]{amsart}
\usepackage{amsmath}

\newtheorem{proposition}{Proposition}
\newtheorem{theorem}{Theorem}

\newcommand{\Z}{\mathbb Z}

\newcommand{\N}{\mathbb N}
\newcommand{\C}{\mathbb C}
\newcommand{\R}{\mathbb R}
\newcommand{\bo}{\mathcal B([0,2\pi))}
\newcommand{\fii}{\varphi}
\newcommand{\hi}{\mathcal H}
\def\<{\langle}
\def\>{\rangle}
\def\d{{\mathrm d}}
\newcommand{\cal}{\mathcal}
\newcommand{\lh}{{\cal L}({\cal H})}
\newcommand{\thi}{{\cal T}({\cal H})}
\newcommand{\st}{{\cal T}({\cal H})^+_1}
\newcommand{\ket}[1]{\left|#1\right\rangle}  
\newcommand{\bra}[1]{\left\langle#1\right|}  
\newcommand{\kb}[2]{\left|#1\right\rangle\left\langle#2\right|}

\begin{document}

\sloppy

\title{On the structure of covariant phase observables}
\author{Juha-Pekka Pellonp\"a\"a}
\address{Department of Physics, University of Turku, 
20014 Turku, Finland}
\email{juhpello@utu.fi}
\maketitle

\begin{abstract}
We study the mathematical structure of covariant phase observables.
Such observables can alternatively be expressed as phase matrices,
as sequences of unit vectors, as sequences of phase states, or as equivalence
classes of covariant trace-preserving operations.
Covariant generalized operator measures are defined by structure matrices 
which form a $\rm W^*$-algebra with phase matrices as its subset.
The properties of the Radon-Nikod\'ym derivatives of phase probability measures 
are studied.
\end{abstract}

\section{Introduction}

Covariant phase observables constitute a particular solution
to the problem of quantum phase (see, e.g.\ \cite{Ho,LaPe1}).
In this paper, we study general mathematical properties of 
covariant phase observables and represent them as covariant trace-preserving
operations (Sec.\ 3). We also analyze the structure matrix $\rm W^*$-algebra 
of covariant generalized operator measures (Sec.\ 4) and 
the pointwise convergence of phase probability densities (Sec.\ 5)

Let $\hi$ be a complex Hilbert space with
a fixed basis $\{\ket n\in\hi\,|\,n\in\N\}$.
Define the number operator $N:=\sum_{n=0}^\infty n\kb n n$ with its usual domain
$\cal D(N):=\left\{\psi\in\hi\,\big|\,n^2|\<n|\psi\>|^2<\infty\right\}$
and the phase shifter $R(\theta):=e^{i\theta N}$ for all $\theta\in\mathbb R$.
Let $\lh$, $\thi$, and $\st$ denote the sets of bounded operators,
trace-class operators, and states (positive trace-one operators) on $\hi$, respectively.

Let $\bo$ denote the $\sigma$-algebra of the Borel subsets of $[0,2\pi)$, and
consider an operator measure $E:\,\bo\to\mathcal L(\mathcal H)$. 
The measure $E$ is normalized if $E([0,2\pi))=I$, positive if
$E(X)\ge O$ for all $X\in\bo$, and phase shift covariant if
$R(\theta)E(X)R(\theta)^*=E(X\oplus\theta)$
for all $X\in\bo$ and for all $\theta\in[0,2\pi)$, where 
$X\oplus\theta:=\{x\in[0,2\pi)\,|\,(x-\theta)({\rm mod}\;2\pi)\in X\}$.
A phase shift covariant normalized positive operator measure is called 
a {\it (covariant) phase observable}.

In the next section we collect some known properties of covariant phase observables.
The new results are contained in Sections 3-5.

\section{The structure of phase observables}
Any covariant phase observable is of the (weakly convergent) form
\begin{equation}\label{phase}
E(X)=\sum_{n,m=0}^\infty c_{n,m}i_{n-m}(X)\kb n m ,\;\;\;\;X\in\bo,
\end{equation}
where $i_k(X):=(2\pi)^{-1}\int_Xe^{ik\theta}\d\theta$ for all $k\in\Z$, and
where the phase matrix $(c_{n,m})_{n,m\in\N}$ 
is a positive semidefinite (complex) matrix with $c_{n,n}=1$, $n\in\N$
(see, e.g.\ Phase Theorem 2.2 of \cite{LaPe1}).
A complex matrix $(c_{n,m})$ is a phase matrix if and only if 
there exist a sequence $(\psi_n)_{n\in\N}$ of unit vectors such that
$c_{n,m}=\<\psi_n|\psi_m\>$, $n,\,m\in\N$ \cite{CaViLaPe}.
A constant sequence,
e.g.\ $\psi_n=|0\>$, $n\in\N$, defines the {\it canonical phase observable}
$$
E_{\rm can}(X):=\sum_{n,m=0}^\infty i_{n-m}(X)\kb n m,
\;\;\;\;X\in\bo,
$$
whereas any orthonormal sequence, e.g.\ $\psi_n=|n\>$, $n\in\N$, 
gives the {\it trivial phase observable}
$$
E_{\rm triv}(X):=i_0(X)\,I,\;\;\;\;X\in\bo.
$$
Next we show how any phase observable can be constructed by using a sequence
of phase states (Theorem \ref{T2}).

Define $\hi_1:=\left\{\psi\in\hi\,\big|\,\sum_{n=0}^\infty|\<n|\psi\>|<\infty\right\}$.
A phase matrix $(c_{n,m}) $ can be interpreted as a 
phase kernel, that is, a positive
(possible unbounded in the norm of $\hi$) sequilinear form $C:\,\hi_1\times\hi_1\to\C$
defined as
$$
C(\fii,\psi):=\sum_{n,m=0}^\infty c_{n,m}\<\fii|n\>\<m|\psi\>,\;\;\;\;\fii,\,\psi\in\hi_1,
$$
where the sum converges absolutely. Keeping this in mind, we may formally write
$$
C=\sum_{n,m=0}^\infty c_{n,m}|n\>\<m|.
$$
Since $R(\theta)\hi_1=\hi_1$ for all $\theta\in[0,2\pi)$ we can define a
continuous integrable\footnote{
Since for all $\fii$, $\psi\in\hi_1$
the series of continuous integrable functions 
$\sum_{n=0}^s\sum_{m=0}^t 
c_{n,m}e_n\overline{e_m}\<\fii|n\>\<m|\psi\>$, where $e_{n}(\theta)=e^{in\theta}$,
$n\in\N$, $\theta\in\R$, 
converges uniformly on $[0,2\pi]$ when $s,\,t\to\infty$, the function 
$\theta\mapsto C(R(-\theta)\fii,R(-\theta)\psi)$ is continuous and integrable.}
function $[0,2\pi]\to\C$
$$
\theta\mapsto C(R(-\theta)\fii,R(-\theta)\psi)
=\sum_{n,m=0}^\infty c_{n,m}e^{i(n-m)\theta}\<\fii|n\>\<m|\psi\>
$$ 
for all $\fii,\,\psi\in\hi_1$, and thus a bounded positive 
sesquilinear form $\hi_1\times\hi_1\to\C$
\begin{eqnarray*}
(\fii,\psi)\mapsto E(X)_{\fii,\psi}&:=&\frac{1}{2\pi}\int_X C(R(-\theta)\fii,R(-\theta)\psi)\d\theta\\
&=&\sum_{n,m=0}^\infty c_{n,m}i_{n-m}(X)\<\fii|n\>\<m|\psi\>
\end{eqnarray*}
for all $X\in\bo$.
The form $(\fii,\psi)\mapsto E(X)_{\fii,\psi}$
has a unique bounded positive extension to $\hi\times\hi$
which is determined by a unique bounded operator, say, $E(X)\in\lh$.
Operators $E(X)$, $X\in\bo$, constitute a covariant phase observable.
The following route to define a phase observable is thus justified:
\begin{enumerate}
\item take a phase matrix $(c_{n,m}) $ and define the
phase kernel $\sum_{n,m=0}^\infty c_{n,m}\kb n m$;
\item act on it by $R(\theta)$ to get 
$$
R(\theta)\sum_{n,m=0}^\infty c_{n,m}\kb n m R(\theta)^*
=\sum_{n,m=0}^\infty c_{n,m}e^{i(n-m)\theta}\kb n m;
$$
\item integrate it over $X\in\bo$ to get a bounded sesquilinear form
$\hi_1\times\hi_1\to\C$,
$$
\frac{1}{2\pi}\int_X R(\theta)\sum_{n,m=0}^\infty c_{n,m}\kb n m R(\theta)^*\d\theta
=\sum_{n,m=0}^\infty c_{n,m} i_{n-m}(X)\kb n m;
$$
\item this has a unique bounded extension $\hi\times\hi\to\C$
which defines the phase observable
$$
E(X):=\sum_{n,m=0}^\infty c_{n,m} i_{n-m}(X)\kb n m.
$$
\end{enumerate}

Let $\hi_\infty$ be a complex Banach space\footnote{
Equip $\hi_1$ with the norm $\psi\mapsto\|\psi\|_1:=\sum_{n=0}^\infty|\<n|\psi\>|$.
The continuous linear mappings $\hi_1\to\C$ form a topological dual $\hi_1'$ of
$\hi_1$. Using the Dirac notation, an element $(F|\in\hi_1'$ can be represented
in the form $(F|=\sum_{n=0}^\infty f_n\<n|$
where $(f_n)_{n\in\N}\subset\C$ and $\sup\left\{|f_n|\,\big|\,n\in\N\right\}<\infty$.
Defining a conjugate form $|F)$ of $(F|$ as a mapping 
$\hi_1\ni\psi\mapsto\overline{(F\ket\psi}\in\C$
we may define the linear space $\hi_\infty$ of conjugate forms of the elements of 
$\hi_1'$.
Thus, using the Dirac formalism, we may write an element $|G)\in\hi_\infty$ of the form
$|G)=\sum_{n=0}^\infty g_n\ket n$
where $(g_n)_{n\in\N}\subset\C$ and $\sup\left\{|g_n|\,\big|\,n\in\N\right\}<\infty$.
We can define the following norm in $\mathcal H_\infty$: 
$|G)\mapsto\left\||G)\right\|_\infty:=\sup\left\{|g_n|\,\big|\,n\in\N\right\}.$
}
of vectors
$\sum_{n=0}^\infty g_n\ket n$ for which the norm 
$\left\|\sum_{n=0}^\infty g_n\ket n\right\|_\infty:=
\sup\left\{|g_n|\,\big|\,n\in\N\right\}<\infty$.
Embedding $\hi$ in $\hi_\infty$ we get the following triplet
$$
\hi_1\subset\hi\subset\hi_\infty.
$$
We have the following theorem \cite{LaPe3}:
\begin{theorem}\label{2}
For any phase matrix $(c_{n,m})$ 
$$
\sum_{n,m=0}^\infty c_{n,m}\<\varphi|n\>\<m|\psi\>=
\sum_{k=0}^\infty\<\varphi|F_k)(F_k|\psi\>,
$$
for all $\varphi$, $\psi\in\hi_1$, that is, briefly,
$$
\sum_{n,m=0}^\infty c_{n,m}\kb n m
=\sum_{k=0}^\infty|F_k)(F_k|
$$
where $|F_k)\in\hi_\infty$ for all $k\in\N$ and $\sum_{k=0}^\infty|\<n|F_k)|^2=1$
for all $n\in\N$.
Conversely, if $\left(|F_k)\right)_{k\in\N}\subset\hi_\infty$
is such that $\sum_{k=0}^\infty|\<n|F_k)|^2=1$ then
$\sum_{k=0}^\infty|F_k)(F_k|$ is a phase kernel.
\end{theorem}
Let $|F)\in\mathcal H_\infty$ and define $|F;\theta):=R(\theta)|F)$ 
and $(F;\theta|:=(F|R(\theta)^*$ for all $\theta\in\R$.
Since $R(\theta')|F;\theta)=|F;\theta+\theta')$ we say that
$|F;\theta)$ is a {\it phase state}.
It easy to see that the following sesquilinear form $\hi_1\times\hi_1\to\C$,
$$
(\fii,\psi)\mapsto\frac{1}{2\pi}\int_X\<\fii|F;\theta)(F;\theta|\psi\>\d\theta,
$$
is positive and bounded for all $X\in\bo$ and it defines a covariant
positive operator measure
\begin{equation}\label{y}
\bo\ni X\mapsto
E_F(X)=\sum_{n,m=0}^\infty\<n|F)(F|m\>i_{n-m}(X)\kb n m\in\lh.
\end{equation}
The operator measure $E_F$ is normalized, that is, a phase observable, if and only if
$|\<n|F)|=1$ for all $n\in\N$, that is, when
$$
|F)=\sum_{n=0}^\infty e^{i\upsilon_n}|n\>
$$
where $(\upsilon_n)_{n\in\N}\subset[0,2\pi)$.
Let $U:=\sum_{n=0}^\infty{e^{i\upsilon_n}}\kb n n$.
Then $E_F$ is a phase observable if and only if
$$
E_F(X)= UE_{\rm can}(X)U^*,\;\;\;\;X\in\bo.
$$
If, for two phase obsevables $E_1$ and $E_2$, the condition
$E_1(X)= UE_2(X)U^*$, $X\in\bo$, holds, we say that $E_1$ is $E_2$ up to unitary 
equivalence, or briefly, $E_1$ is $E_2$ (u.e.).
Thus, using Theorem \ref{2} we get a variant of Phase Theorem 2.2 of \cite{LaPe1}:
\begin{theorem}\label{T2}
$E$ is a phase observable if and only if for all $X\in\bo$
$$
E(X)=\mathop{\hbox{\rm w-lim}}_{n\to\infty}\sum_{k=0}^n E_{F_k}(X)
$$
where $E_{F_k}(X)$ is the bounded operator defined by a sesquilinear form
$$
\frac{1}{2\pi}\int_X|F_k;\theta)(F_k;\theta|\d\theta
$$
where $|F_k)\in\hi_\infty$, $k\in\N$, and
$\sum_{k=0}^\infty|\<n|F_k)|^2=1$.

The phase observable $E$ is defined by a single phase state 
if and only if $E$ is $E_{\rm can}$ (u.e.).
\end{theorem}
Since the sequence $n\mapsto \sum_{k=0}^n E_{F_k}(X)$ is increasing
$E(X)=\mathop{\hbox{\rm s-lim}}_{n\to\infty}\sum_{k=0}^n E_{F_k}(X)$ also.

\section{Phase observables as operations}
A linear mapping $\Phi:\thi\to\thi$ is a {\it covariant trace-preserving operation} 
if it is covariant ($R(\theta)\Phi(T)R(\theta)^*=\Phi(R(\theta)TR(\theta)^*)$, 
$\theta\in[0,2\pi)$, $T\in\thi$),
trace-preserving ($\mathrm{tr}(\Phi(T))=\mathrm{tr}(T)$, $T\in\thi$), and 
positive ($\Phi(\st)\subseteq\st$) (for the theory of operations, see e.g.\ 
\cite{Da, Kr}). 
We prove next a theorem essentially due to
Hall and Fuss \cite{HaFu,Ha}.
\begin{theorem}\label{T3}
A mapping $E:\,\bo\to\lh$ is a phase observable if and only if
\begin{equation}\label{i}
{\rm tr}(TE(X))={\rm tr}(\Phi(T)E_{\rm can}(X))
\end{equation}
for all $X\in\bo$ and $T\in\thi$ where 
$\Phi:\thi\to\thi$ is a covariant trace-preserving operation.
\end{theorem}
\begin{proof}
Let $E$ be a phase observable with the phase matrix $(c_{n,m})$.
For all $T\in\thi$ define
\begin{equation}\label{ii}
\Theta(T):=\sum_{n,m=0}^\infty c_{m,n}T_{n,m}\kb n m.
\end{equation}
Since $T=\alpha T_\alpha-\beta T_\beta+i\gamma T_\gamma-i\delta T_\delta$ where
$T_\alpha$, $T_\beta$, $T_\gamma$, and $T_\delta$ 
are states, and
$\alpha$, $\beta$, $\gamma$, and $\delta$ are nonnegative real numbers, it suffices
to consider only states. Thus, assume that $T$ is a state.
Since $\sup\left\{|\<\fii|\Theta(T)\psi\>|\,\big|\,\|\fii\|\le1,\,\|\psi\|\le1\right\}\le
\sup\left\{\sum_{n,m=0}^\infty|T_{n,m}|\,|\<\fii|n\>|\,|\<m|\psi\>|\;\Big|\;\|\fii\|\le1,
\,\|\psi\|\le1\right\}\le1$ it follows that $\Theta(T)$ is a bounded operator. 
Using a decomposition $T=\sum_{j=0}^\infty|\phi_j\>\<\phi_j|$,
$\phi_j\in\hi$, $j\in\N$,
one sees that $\<\psi|\Theta(T)\psi\>=\sum_{j=0}^\infty
\sum_{n,m=0}^\infty\overline{\<m|\phi_j\>\<\psi|m\>}
c_{m,n}\<n|\phi_j\>\<\psi|n\>\ge0$ for all $\psi\in\hi_1$ and, thus, $\Theta$ is positive. 
Since $\sum_{n=0}^\infty\<n|\Theta(T)|n\>=1$, 
$\Theta(T)$ is a trace-one operator. Moreover,
${\rm tr}(TE(X))={\rm tr}(\Theta(T)E_{\rm can}(X))$, $X\in\bo$, and $\Theta$ is covariant.
Thus, $\Theta$ is a covariant trace-preserving operation. 
The converse part is trivial.
\end{proof}

There are many covariant trace-preserving operations $\Phi$ which satisfy Equation
(\ref{i}) for a given $E$. One such operation $\Theta$ is defined in (\ref{ii}).
It is the identity operation in the case of the canonical phase
whereas for the trivial phase it is of the form
$\Theta(T)=\sum_{n=0}^\infty T_{n,n}\kb n n$.
We note also that, in the case of the trivial phase, $T\mapsto T_{0,0}\kb 1 1+T_{1,1}\kb 0 0+
\sum_{n=2}^\infty T_{n,n}\kb n n$ is an other operation fulfilling Theorem \ref{T3}.
Since the diagonal
elements $T_{n,n}$ do not "contain" any phase information of the state $T$ we see
that the trivial phase "loses" all phase information. 
In the general case, if $c_{n,m}=0$ for some $n\ne m$, there are vector states
(other than number states) $\psi:=d_n\ket n + d_m\ket m$, $d_n,\,d_m\in\C\setminus\{0\}$,
$|d_n|^2+|d_m|^2=1$, for which the probability measure $X\mapsto\<\psi|E(X)\psi\>$
is random. Next we study the properties of $\Theta$.

Let $E$ be a phase observable with the phase matrix $(c_{n,m})$, and let 
$\Theta(T)=\sum_{n,m=0}^\infty c_{m,n}T_{n,m}\kb n m$ for all $T\in\thi$.
The dual mapping $\Theta^*:\,\lh\to\lh$ of an operation $\Theta$ defined by
the relation ${\rm tr}(T\Theta^*(A))={\rm tr}(\Theta(T)A)$, $A\in\lh$, $T\in\thi$, 
is a positive linear mapping, and
$$
\Theta^*(A)=\sum_{n,m=0}^\infty c_{n,m}A_{n,m}\kb n m,\;\;\;\;A\in\lh.
$$
From Theorem \ref{2} one gets (weakly)
$$
\Theta(T)=\sum_{k=0}^\infty A_k T A_k^*,\;\;\;\;T\in\thi,
$$
where $A_k:=\sum_{n=0}^\infty(F_k|n\>\kb n n$ for all $k\in\N$ showing that
$\Theta$ is {\it completely positive} (see the First Representation Theorem
of \cite{Kr}). Note that $\sum_{k=0}^\infty A_k A_k^*=I$ and
$\Theta^*(A)=\sum_{k=0}^\infty A_k^* A A_k$, $A\in\lh$.

Let $\Theta^+_1:\,\st\to\st$ be the restriction of $\Theta$ to the set of states.
\begin{theorem}
\begin{enumerate}
\item $\Theta$ and $\Theta^+_1$ are injections if and only if $c_{n,m}\ne0$ for all $n,\,m\in\N$;
\item $\Theta^+_1$ is surjection if and only if $E$ is $E_{\rm can}$ (u.e.);
\item $\Theta^+_1$ is bijection if and only if $E$ is $E_{\rm can}$ (u.e.);
\item $\Theta$ preserves pure states ($\Theta(\kb\psi\psi)^2=\Theta(\kb\psi\psi)$ for all
unit vectors $\psi\in\hi$) if and only if $E$ is $E_{\rm can}$ (u.e.).
\end{enumerate}
\end{theorem}
\begin{proof}
It is easy to see that $\Theta$ and $\Theta^+_1$ are injections if and only if
$c_{m,n}T_{n,m}=0$ for all $n,\,m\in\N$ where $T\in\thi$ implies that $T=O$.
Thus, $\Theta$ and $\Theta^+_1$ are injections if and only if $c_{n,m}\ne0$ for all $n,\,m$.

Suppose that $\Theta^+_1$ is surjection. If $c_{m,n}=0=c_{n,m}$ 
for some $n\ne m$ then
$\Theta^+_1(T)\ne T':=(\ket n+\ket m)(\bra n+\bra m)/2$ for all $T\in\st$ and, thus, $c_{n,m}\ne0$ for all $n,\,m$ and
$\Theta^+_1$ is injection and bijection. 
If $|c_{n,m}|<1$ for some $n\ne m$ then
there is no state $T$ such that $\Theta^+_1(T)=T'$. Thus, $|c_{n,m}|=1$, $n,\,m\in\N$, and
$E=E_{\rm can}$ (u.e.). 
This proves items (2) and (3).

Let $\psi:=\sum_{n=0}^\infty d_n\ket n$ where $d_n>0$ for all $n$ and 
$\sum_{n=0}^\infty d_n^2=1$.
Now $\Theta(\kb\psi\psi)^2=\Theta(\kb\psi\psi)$ implies that
$\sum_{n=0}^\infty|c_{n,m}|^2 d_n^2=1$ for all $m$ which shows that
$|c_{n,m}|=1$, $n,\,m\in\N$, and $E=E_{\rm can}$ (u.e.). 
This completes the proof.
\end{proof}

\section{Covariant GOMs and phase matrices}
The standard way to represent an observable in quantum mechanics is to find an 
appropriate self-adjoint operator, or an idempotent POM, 
which describes that observable. However, in many cases this representation is too narrow and 
it is convenient to give up the idempotency (see, e.g.\ \cite{OQP}).
The strength of POMs is that they associate a probability measure to {\it all} states.
If we restrict ourselves to a subset of (vector) states
to be called physical states we can give up the positivity of POM and require
that the operator measure gives a probability measure (via trace formula) only for
physical states. Actually, we do not have to assume that the observable can even 
be "defined" for other states that physical ones. 
Hence, define a set of physical states $\cal V$. It is a linear subspace of the Hilbert 
space of the physical system. The linearity is assumed because of the possibility
to superpose the physical states. Let $\cal{SL(V,V;\C)}$ be the set of sesquilinear
forms from $\cal V\times\cal V$ to $\C$ (the first argument is antilinear).
A {\it generalized operator measure} \cite{Pe}, or a {\it GOM}, $G$ is the mapping from 
the $\sigma$-algebra $\cal A$ of the set of measurement outcomes $\Omega$ 
to $\cal{SL(V,V;\C)}$ such that $\cal A\ni X\to[G(X)](\fii,\psi)\in\C$ is a complex measure
for all $\fii,\,\psi\in\cal V$. 
It is {\it normalized} if $[G(\Omega)](\fii,\psi)=\<\fii|\psi\>$, $\fii,\,\psi\in\cal V$. 

In the case of phase, it is natural to assume that $\Omega=[0,2\pi)$,
$\cal A=\bo$, and $\cal V$ contains number states, coherent states, etc.
Since they are elements of $\cal H_1$ we assume that $\cal V=\cal H_1$.
If we study the coherent state phase measurements with the associated 
GOM $E:\,\bo\to\cal{SL}(\cal H_1,\cal H_1;\C)$,
it is natural to assume the following phase shift covariance condition:
\begin{equation}\label{kova}
[E(X)](|ze^{-\alpha}\>,|ze^{-\alpha}\>)=[E(X\oplus\alpha)](|z\>,|z\>)
\end{equation}
for all $X\in\bo$, $z\in\C$, and $\alpha\in[0,2\pi)$.
The following GOMs are solutions of (\ref{kova}):
\begin{equation}\label{nm}
[E(X)](\fii,\psi)=\sum_{n,m=0}^\infty d_{n,m}i_{n-m}(X)\<\fii|n\>\<m|\psi\>
\end{equation}
where 
$(d_{n,m})\in\C^{\N\times\N}$, $\sup\left\{|d_{n,m}|\,\big|\,n,m\in\N\right\}<\infty$,
$X\in\bo$, and $\fii,\,\psi\in\cal H_1$.
We use the following short notation for $E$:
$$
E(X)=\sum_{n,m=0}^\infty d_{n,m}i_{n-m}(X)|n\>\<m|,
$$
and we say that $E$ is a {\it covariant GOM}
defined by the {\it structure matrix} $(d_{n,m})$.
Note that $E([0,2\pi))=\sum_{n=0}^\infty d_{n,n}|n\>\<n|$ can be extended to a unique 
bounded operator.
If $d_{n,n}=1$, $n\in\N$, then $E$ is normalized.
If $(d_{n,m})$ is a phase matrix then $E$ is a phase observable.
For all $\fii,\,\psi\in\cal H_1$, the complex measure $X\to[E(X)](\fii,\psi)$
has a continuous density which is 
$$
\theta\mapsto\sum_{n,m=0}^\infty d_{n,m}e^{i(n-m)\theta}\<\fii|n\>\<m|\psi\>.
$$

Let $\cal M_\infty$ be a set of structure matrices
$(d_{n,m})_{n,m\in\N}\in\C^{\N\times\N}$, 
$\sup\left\{|d_{n,m}|\,\big|\,n,m\in\N\right\}<\infty$.
Since for all $(d_{n,m})\in\cal M_\infty$ we have a unique 
covariant genaralized operator 
measure $E$ defined in (\ref{nm}), we can identify $(d_{n,m})$ with $E$.
Now $\cal M_\infty$ is a $\rm W^*$-algebra (over $\C$)
with the norm $\|(d_{n,m})\|:=\sup\left\{|d_{n,m}|\,\big|\,n,m\in\N\right\}<\infty$.
The summation, scalar product, and algebra product are
defined pointwise. 
Let $\star$ be the algebra product operation, that is,
$(d_{n,m})\star(e_{n,m}):=(d_{n,m}e_{n,m})$.
The identity of $\cal M_\infty$ is the canonical phase matrix
$(c_{n,m})$ with $c_{n,m}=1$, $n,\,m\in\N$.
The algebra $\cal M_\infty$ is commutative and 
the involution is $(d_{n,m})\mapsto(d_{n,m})^*:=(\overline{d_{n,m}})$.
The unique pre-dual of $\cal M_\infty$ is the Banach space $\cal M_1$ of 
matrices $(d_{n,m})$ for which $\sum_{n,m=0}^\infty|d_{n,m}|<\infty$. 
A matrix $(d_{n,m})\in\cal M_\infty$ has an inverse if and only if $d_{n,m}\ne0$
for all $n,\,m\in\N$. The inverse is $(d_{n,m}^{-1})$.
A matrix $(d_{n,m})\in\cal M_\infty$ is positive if $d_{n,m}\ge0$ for all $n,\,m\in\N$.
However, we are not interested in this standard positivity;
we rather study positive semidefinitess of matrices.

The positive semidefinite matrices of $\cal M_\infty$ form a convex cone. 
We denote it by $\cal M_\infty^+$. 
Any $(d_{n,m})\in\cal M_\infty^+$ defines a covariant positive operator measure $E$
via Equation (\ref{nm}).
The phase matrices are such matrices of $\cal M_\infty^+$
whose diagonal elements equal one. Let $\cal{C}$ be the $\sigma$-convex\footnote{
$\sigma$-convex means that for any sequence of phase matrices $(C_k)_{k\in\N}$ and
for any sequence of nonnegative real numbers $(\lambda_k)_{k\in\N}$ for which
$\sum_{k=0}^\infty\lambda_k=1$ the series $n\mapsto\sum_{k=0}^n\lambda_kC_k$ 
converges to a phase matrix (with respect to the norm of $\cal M_\infty$)}
set of phase matrices. Phase matrices define phase observables.
The phase matrices of phase observables unitarily equivalent to $E_{\rm can}$
are only phase matrices which have phase matrix inverses. 
Note that $\cal C^*=\cal C$, and
for all $(c_{n,m})\in\cal C$ the norm $\|(c_{n,m})\|=1$, that is, 
all phase matrices lie on the unit ball.

We can embed the bounded operators, trace-class operators, and states in $\cal M_\infty$.
We simply define 
\begin{eqnarray*}
\cal L&:=&\left\{(A_{n,m})\in\cal M_\infty\,\Big|\,\sum_{n,m=0}^\infty A_{n,m}\kb n m\in\lh\right\},\\
\cal T&:=&\left\{(T_{n,m})\in\cal M_\infty\,\Big|\,\sum_{n,m=0}^\infty T_{n,m}\kb n m\in\thi\right\},\\
\cal T^+_1&:=&\left\{(T_{n,m})\in\cal M_\infty\,\Big|\,\sum_{n,m=0}^\infty T_{n,m}\kb n m\in\st\right\}.
\end{eqnarray*}
Thus, $\cal T^+_1$ contains such $(T_{n,m})\in\cal M_\infty^+$ for which
$\sum_{n=0}^\infty T_{n,n}=1$. 
Note that 
$\cal T\cap\cal C=\emptyset$ and $\cal C\not\subseteq\cal L\ne\cal M_\infty$.

As we saw in the previous section,
for any phase matrix $(c_{n,m})$ and a state $(T_{n,m})$ the product
$(c_{n,m})\star(T_{n,m})$ is a state. Thus, $\cal C\star\cal T^+_1=\cal T^+_1$.
An operation $\Theta$ defined in (\ref{ii}) corresponds a 
mapping 
$\cal T\ni(T_{n,m})\mapsto(c_{m,n})\star(T_{n,m})\in\cal T$, $(c_{n,m})\in\cal C$,
which is continuous with respect to the trace-norm.
If $\Theta_1$ and $\Theta_2$ are
the operations (defined in (\ref{ii})) 
of phase observables $E_1$ and $E_2$ with $(c^1_{n,m})$ 
and $(c^2_{n,m})$, respectively, then the matrix $(c^1_{m,n})\star(c^2_{m,n})$ 
corresponds the composition operation
$\Theta_1\circ\Theta_2$.
Note that $\Theta_1\circ\Theta_2=\Theta_2\circ\Theta_1$.

Let $(d_{n,m})$ and $(e_{n,m})$ be elements of $\cal M_\infty^+$.
Now there exist vector sequences $(\fii_n)_{n\in\N}$ and $(\psi_n)_{n\in\N}$
such that $d_{n,m}=\<\fii_n|\fii_m\>$ and $e_{n,m}=\<\psi_n|\psi_m\>$
for all $n,\,m\in\N$ \cite{CaViLaPe}.
Now $d_{n,m}e_{n,m}=\<\fii_n\otimes\psi_n|\fii_m\otimes\psi_m\>$, $n,\,m\in\N$, and
$(d_{n,m})\star(e_{n,m})$ is positive semidefinite.\footnote{
For any sequence $(f_n)_{n\in\N}\subset\C$ for which
$f_n\ne0$ for only finite many $n\in\N$ the sum
$\sum_{n,m=0}^\infty\overline{f_n}d_{n,m}e_{n,m}f_m=
\left\|\sum_{n=0}^\infty f_n|\fii_n\otimes\psi_n\>\right\|^2\ge0$.
This shows that $(d_{n,m}e_{n,m})$ is positive semidefinite.}
Hence, $\cal M_\infty^+\star\cal M_\infty^+=\cal M_\infty^+$ and
$\cal C\star\cal C=\cal C$.

Let $(d_{n,m})\in\cal M_\infty^+$. Now we can write $d_{n,m}=
\sum_{k=0}^\infty d^{(k)}_{n,m}$ where $d^{(k)}_{n,m}=\<n|F_k)(F_k|m\>$
for all $n,\,m\in\N$, and $|F_k)\in\cal H_\infty$, $k\in\N$. Hence, the finite sums of 
matrices $\left(\<n|F)(F|m\>\right)_{n,m\in\N}$, 
$|F)\in\cal H_\infty$, form a dense subset of $\cal M_\infty^+$. 
Every $|F)\in\cal H_\infty$ defines 
a covariant positive operator measure $E_F$ of Equation (\ref{y}).

Following \cite{Ha}, we can define a certain ordering relation 
on $\cal M_\infty$ as follows:
$(d_{n,m})\preceq(e_{n,m})$ if $(d_{n,m})=(e_{n,m})\star(f_{n,m})$ for some 
$(f_{n,m})\in\cal M_\infty$. 
Let $(1)_{n,m\in\N}$ and $(\delta_{n,m})_{n,m\in\N}$ be the phase matrices of 
the canonical and the trivial phase observables, respectively.
Now $(d_{n,m})_{n,m\in\N}\preceq(1)_{n,m\in\N}$ for all $(d_{n,m})\in\cal M_\infty$ and
$(\delta_{n,m})\preceq(c_{n,m})$ for all $(c_{n,m})\in\cal C$.
Note that $\preceq$ is not a partial ordering. It does not satisfy the antisymmetry
condition. 

Define the following equivalence relation in $\cal C$:
$$
(c_{n,m})\simeq(d_{n,m})\hbox{ if }(c_{n,m})=(d_{n,m})\star\left(e^{i(\upsilon_n-\upsilon_m)}\right),
\;\;\;\;(\upsilon_n)_{n\in\N}\subset[0,2\pi).
$$
Denote the equivalence class of $(c_{n,m})\in\cal C$ by $[(c_{n,m})]$, and
define a partial ordering $\preceq$ in the set of equivalence classes as follows:
$[(c_{n,m})]\preceq[(d_{n,m})]$ if $(c_{n,m})=(d_{n,m})\star(e_{n,m})$ for some 
$(e_{n,m})\in\cal C$. 
Now $[(\delta_{n,m})]\preceq[(c_{n,m})]\preceq[(1)]$ for all $(c_{n,m})\in\cal C$ and,
thus, the equivalence class of the canonical phase matrix is the upper bound.


\section{On the pointwise convergence of phase kernels}
As we have seen, a phase observable $E$ is determined uniquely
by a phase matrix $(c_{n,m})$ via Equation (\ref{phase}).
For any trace-class operator $T$ we can define a complex measure
$X\mapsto p^E_T(X):={\rm tr}(TE(X))$ which is absolutely continuous with respect to
the normalised Lebesgue measure and, thus, has a Radon-Nikod\'ym derivative
$g^E_T$ such that $p^E_T(X)= (2\pi)^{-1}\int_X g^E_T(\theta)\d\theta$, $X\in\bo$.
Following Equation (\ref{phase})
it is tempting to write 
$g^E_T(\theta)=\sum_{n,m=0}^\infty T_{m,n}c_{n,m}e^{i(n-m)\theta}$ 
where the summation converges pointwise 
for $\d\theta$-almost all $\theta\in\R$. 
But is it possible? In this section we study this
problem.

Let us start with the simplest case. Let $E$ be the canonical phase, and let
$T=\kb\fii\psi$ where $\fii,\,\psi\in\hi$.
From Carleson Theorem \cite{Ca}
we know that any $L^2$-Fourier series converges pointwise
for almost all $\theta\in\R$. Thus, we get
$$
g^{E_{\rm can}}_{\kb\fii\psi}(\theta)=\sum_{n=0}^\infty\<\psi|n\>e^{in\theta}
\sum_{m=0}^\infty\<m|\fii\>e^{-im\theta}
=\sum_{n,m=0}^\infty\<m|\fii\>\<\psi|n\>e^{i(n-m)\theta}
$$
for almost all $\theta\in\R$.
Let then $T$ be an arbitrary trace-class operator, and let $E$ be any phase observable
with the covariant trace-preserving operation $\Phi$ of Theorem \ref{T3}.
Now we can write $\Phi(T)=T_\alpha-T_\beta+iT_\gamma-iT_\delta$ where 
the operators $T_u$ are positive trace-class operators
with decompositions 
$T_u=\sum_{k=0}^\infty|\fii^{(u)}_k\>\<\fii^{(u)}_k|$,
$\fii^{(u)}_k\in\hi$, $k\in\N$, where 
$u=\alpha,\,\beta,\,\gamma,\,\delta$.
Thus,
$$
g^E_T(\theta)=g^{E_{\rm can}}_{T_\alpha}(\theta)-g^{E_{\rm can}}_{T_\beta}(\theta)+ig^{E_{\rm can}}_{T_\gamma}(\theta)
-ig^{E_{\rm can}}_{T_\delta}(\theta)
$$
and, by monotonic convergence,
\begin{equation}\label{n}
g^{E_{\rm can}}_{T_u}(\theta)=\sum_{k=0}^\infty
\sum_{n,m=0}^\infty\<m|\fii^{(u)}_k\>\<\fii^{(u)}_k|n\>e^{i(n-m)\theta}
\end{equation}
for all $u=\alpha,\,\beta,\,\gamma,\,\delta$ and for almost all $\theta\in\R$.
We will get a similar equation without using the operation $\Phi$.
Namely, by using Theorem \ref{T2} one gets for any $\kb\fii\fii$
\begin{equation}\label{m}
g^E_{\kb\fii\fii}(\theta)=\sum_{k=0}^\infty\sum_{n,m=0}^\infty\<\fii|n\>\<n|F_k)(F_k|m\>
\<m|\fii\>e^{i(n-m)\theta}
\end{equation}
for almost all $\theta\in\R$.
A problem of Equations (\ref{n}) and (\ref{m})
is that it is not clear that we can change the order of $k$- and $(n,m)$-sums.
So we have to consider other methods.

First we prove a simple proposition.
Let $B$ be a complex Banach space, and
let $S:\,\hi\times\hi\to B$ be a bounded sesquilinear form 
(the first argument is antilinear), that is,
$\|S\| :=\sup\left\{\|S(\fii,\psi)\|\,\big|\,\|\fii\|\le1,\,\|\psi\|\le1\right\}<\infty$.
Note that $\thi$ is equipped with the trace norm.
\begin{proposition}\label{p1}
Denote $S_{n,m}:=S(\ket n,\ket m)$ for all $n,\,m\in\N$. Then for all $\fii,\,\psi\in\hi$
$$
S(\fii,\psi)=\lim_{s,t\to\infty}\sum_{n=0}^s\sum_{m=0}^t S_{n,m}
\<\fii|n\>\<m|\psi\>,
$$
that is,
$$
S=\sum_{n,m=0}^\infty S_{n,m}|n\>\<m|
$$
weakly, and $S$ can be uniquely extended to
a continuous linear mapping $\tilde S:\,\thi\to B$,
$$
T\mapsto\tilde S(T):=\sum_{n,m=0}^\infty S_{n,m}T_{m,n}:=\lim_{s,t\to\infty}
\sum_{n=0}^s\sum_{m=0}^t S_{n,m}T_{m,n}
$$
where $T_{m,n}:=\<m|T|n\>$, $n,\,m\in\N$. Clearly,
$\tilde S(\kb\psi\fii)=S(\fii,\psi)$ for all $\fii,\,\psi\in\hi$.
\end{proposition}

\begin{proof}
For $\fii,\,\psi\in\hi$ one gets
$$
\left\|S(\fii,\psi)-S(P_s\fii,P_t\psi)\right\|
\le \|S\| \,\|\fii\|\,\|\psi-P_t\psi\|+\|S\| \,\|\fii-P_s\fii\|\,\|P_t\psi\|\to0
$$
when $s,\,t\to0$ where $P_s:=\sum_{n=0}^s|n\>\<n|$.
Fix $T\in\st$. One can write $T=\sum_{k=0}^\infty\lambda_k|\fii_k\>\<\fii_k|$ where 
$\lambda_k\in[0,1]$, $\sum_{k=0}^\infty\lambda_k=1$,
$\fii_k\in\hi$, and $\|\fii_k\|=1$ for all $k\in\N$.
Define $\alpha^T:=\sum_{k=0}^\infty\lambda_k S(\fii_k,\fii_k)$ and 
$\alpha^T_{s,t}:=\sum_{k=0}^\infty\lambda_k S(P_s\fii_k,P_t\fii_k)=
\sum_{n=0}^s\sum_{m=0}^t S_{n,m}T_{m,n}$ 
which exist since
$\sum_{k=0}^\infty\lambda_k=1$ and $\|S(\fii,\psi)\|\le \|S\| $ for all vectors $\fii,\,\psi$ 
with $\|\fii\|\le1$, $\|\psi\|\le1$.
Also we see that $\|\alpha^T\|\le\|S\|$.
By the dominated convergence theorem
\begin{eqnarray*}
\|\alpha^T-\alpha^T_{s,t}\|&\le&
\sum_{k=0}^\infty\lambda_k\|S(\fii_k,\fii_k)-S(P_s\fii_k,P_t\fii_k)\|\\
&\le&\|S\| \sum_{k=0}^\infty\lambda_k\left(\|\fii_k-P_t\fii_k\|+\|\fii_k-P_s\fii_k\|\right)\to0
\end{eqnarray*}
when $s,\,t\to\infty$. As we can easily see from the beginning of the proof,
the matrix elements $T_{n,m}$ define the operator $T$ uniquely and, thus,
$\tilde S(T):=\alpha^T$ is well-defined.
Since any $T\in\thi$ can be uniquely written in the form
$T=\alpha T_\alpha-\beta T_\beta+i\gamma T_\gamma-i\delta T_\delta$ where
$T_\alpha$, $T_\beta$, $T_\gamma$, and $T_\delta$ 
are states, and
$\alpha$, $\beta$, $\gamma$, and $\delta$ are nonnegative real numbers
we can define 
$\tilde S(T):=\alpha\tilde S(T_\alpha)-\beta\tilde S(T_\beta)+i\gamma\tilde S(T_\gamma)
-i\delta\tilde S(T_\delta)$.
The rest of the proof follows immediately.
\end{proof}
Note that it follows from Proposition \ref{p1} that any bounded operator $A$
can be written in the form $A=\sum_{n,m=0}^\infty A_{n,m}\kb n m$ (weakly)
and $\mathrm{tr}(AT)=\sum_{n,m=0}^\infty A_{n,m}T_{m,n}$ 
for any $T\in\thi$ where $A_{n,m}:=\<n|A|m\>$, $n,m\in\N$.

Let $E$ be a phase observable with
$(c_{n,m})$, and let $g^E_T$ be a Radon-Nikod\'ym derivative of the 
complex measure $p^E_T$ associated to $T\in\thi$.
The sesquilinear 
mapping $\hi\times\hi\ni(\fii,\psi)\mapsto g^E_{\kb\psi\fii}\in L^1[0,2\pi)$ is
bounded since by using the polarization identity and the parallelogram law
$(2\pi)^{-1}\int_0^{2\pi}\left|g^E_{\kb\psi\fii}(\theta)\right|\d\theta\le\|\psi\|^2+\|\fii\|^2$ 
for all $\psi,\,\fii\in\hi$.
From Proposition \ref{p1} one gets
$$
g^E_T=\sum_{n,m=0}^\infty T_{m,n}c_{n,m}e_n\overline{e_m}
$$
where $e_n(\theta)=e^{in\theta}$ and the double series converges
with respect to the $L^1$-norm. This implies \cite[Theorem 3.12, p.\ 68]{Ru} the following theorem:
\begin{theorem}
There exists a subsequence $\N\ni k\mapsto n_k\in\N$, $n_1<n_2<n_3<...$, such that
$$
g^E_T(\theta)=\lim_{k\to\infty}\sum_{n,m=0}^{n_k}T_{m,n}c_{n,m}e^{i(n-m)\theta}
$$
for almost all $\theta\in\R$.
\end{theorem}

It can be shown \cite[Theorem 7.8, p.\ 140]{Ru} that 
$$
g^E_T(\theta)=2\pi\frac{\mathrm dp^E_T[0,x)}{\mathrm d x}\Big|_{x=\theta}
$$
for almost all $\theta\in\mathbb[0,2\pi)$. 
Thus, by direct calculation one gets
$$
g^E_T(\theta)=\lim_{\epsilon\to0+}
\sum_{n,m=0}^\infty T_{m,n}c_{n,m}e^{i(n-m)\theta}f^{(1)}_{n-m}(\epsilon)
$$
for almost all $\theta\in\R$ 
where $f^{(1)}_{k}(\epsilon):=\left(e^{ik\epsilon}-1\right)/(ik\epsilon)$, $k\ne 0$,
and $f^{(1)}_{0}(\epsilon)=1$. Thus, $\lim_{\epsilon\to0+}f^{(1)}_{k}(\epsilon)=1$ for all 
$k\in\mathbb Z$.
Also, by a theorem of Fatou \cite[p.\ 34]{Hof}, one can show that
\begin{eqnarray*}
g^E_T(\theta)&=&
\lim_{\epsilon\to0+}
\sum_{n,m=0}^\infty T_{m,n}c_{n,m}e^{i(n-m)\theta}f^{(2)}_{n-m}(\epsilon)
\end{eqnarray*}
for almost all $\theta\in\R$
where $f^{(2)}_k(\epsilon):=(1-\epsilon)^{|k|}$ for all $k\in\mathbb Z$
and the double series converges absolutely when $\epsilon\in(0,1]$.
Also $\lim_{\epsilon\to0+}f^{(2)}_k(\epsilon)=1$ for all $k\in\Z$.

Since the operators
$$
C^{(j)}_{\epsilon}:=\sum_{n,m=0}^\infty c_{n,m}f^{(j)}_{n-m}(\epsilon)|n\rangle\langle m|,
\;\;\;j=1,\,2,
$$
are bounded with 
$\|C^{(1)}_{\epsilon}\|\le2\pi/\epsilon$ and 
$\|C^{(2)}_{\epsilon}\|\le2/\epsilon-1$
for each $\epsilon\in(0,1]$ it follows that
$$
g^E_T(\theta)=\lim_{\epsilon\to0+}\hbox{tr}\left[TR(\theta)C^{(j)}_\epsilon R(\theta)^*\right],
\;\;\;j=1,\,2,
$$
for almost all $\theta\in\mathbb R$. 

\subsection*{Acknowledgments.}
The author thanks Dr.\ Pekka J.\ Lahti and Dr.\ Kari Ylinen for many
fruitful discussions.

\end{document}